# Application of Neural Network in Optimization of Chemical Process

Fei Liang, Taowen Zhang

East China University of Science and Technology, 200237 Shanghai, China

**Abstract:** Artificial neural network (ANN) has been widely used due to its strong nonlinear mapping ability, fault tolerance and self-learning ability. This article summarizes the development history of artificial neural networks, introduces three common neural network types, BP neural network, RBF neural network and convolutional neural network, and focuses on the practical application in chemical process optimization, especially the results achieved in multi-objective control optimization and process parameter improvement.

## 1 Introduction

Artificial neural network is a research field involving a wide range of disciplines and strong comprehensiveness. It is attracting wide interest and attention from different industries. The specific forms of neural networks are different, and they are constantly evolving. They can be broadly defined as a complex network system composed of a large number of simple components widely connected. As early as the embryonic period of the Turing machine, a long-term vision of intelligent science being able to think like humans was proposed [1]. After decades, artificial intelligence has developed two classic schools of symbolism and joint conclusions [2]. As a model representative of connectionism, the computational structure and learning rules of artificial neural networks (ANN) are inspired by the highly connected biological neural network information transmission mechanism. ANN has the ability to handle nonlinear problems in parallel, as well as the ability to adapt, self-organize, and self-learn. Neural networks have received extensive attention in recent years, and have achieved a series of successful applications in natural language processing, computer vision, pattern recognition, and intelligent control.

When it comes to the interaction of various experimental factors or even human operations in a specific chemical reaction system, traditional methods make the problem solving complicated. The "black box" nature of neural network makes it a great advantage in solving chemical process problems. Starting from the development process and classification of artificial neural networks, this article mainly introduces the application cases of neural networks in chemical process control and process optimization, and looks forward to the future development prospects.

**2 The development history of artificial neural networks**

The initial research work on neural networks in the last century was mainly divided into three directions: One is to explore the biological structure and mechanism of neural networks in the human brain, which is also the original intention of studying neural networks. The second is the use of microelectronics or optical devices to form a network with certain functions, which is a concern in the field of computer manufacturing. The third is to use artificial neural networks as a means and method to solve problems, which are often difficult or impossible to solve by traditional means. It is the threshold weighted sum model (now called the M-P model) proposed by McCulloch and Pitts in 1943 that really proposed the current neural network model in the usual sense and became a milestone in the development of the entire neural network theory [3]. They determined the output of the neural network as a weighted sum function of the external input signal, and when the weighted sum exceeds the threshold, the system can be activated. The weighting coefficient is determined by the characteristics of the object described, and it can be obtained through step-by-step learning or training. The proposal of the M-P model means that the research of neurodynamic model has entered the first peak period. The research purpose of this period is mainly to find a qualitative or semi-quantitative way of describing brain function, rather than its practical application.

Between 1957 and 1962, Rosenblatt's "hierarchical neural network" [4] theory marked the culmination of the first research climax. But in 1969, Minsky and Papert based on Rosenblatt's discovery, proposed an analysis of the computing power of the perceptron, and wrote the book "*Perceptron*", holding a negative attitude towards the frontiers of neural network research [5]. Although many limitations were overcome later, many artificial neural network research work stopped at that time.

In 1982, Professor Hopfield of the California Institute of Technology proposed a dynamic evolutionary neural network model that is easy to implement with circuits [6], and later named it after him. Although there are still different evaluations of this model until today, it is generally not denied that this theoretical model announces the second peak of artificial neural network research. It is particularly worth pointing out that the research group led by Rumelhart et al. developed the back-propagation learning algorithm relatively well, which has become the most influential algorithm so far [7]. At the same time, the radial basis network (RBF) proposed by Broomhead and Lowe simulates the local response characteristics of neurons so that the network has a fast learning convergence rate [8]. In 1987, the first International Neural Network Conference was held in San Diego. Many well-known experts and scholars gave reports, and the research results were fruitful.

In the 1990s, the development of neural networks reached a new stage. On the one hand, with the improvement of theory and technology, the structure of neural networks became more and more complex and diversified; on the other hand, deep neural networks for practical engineering applications became more and more mature. . Neural network is a multidisciplinary research. The current development of computer hardware, the accumulation of big data and the integration of other knowledge technologies have brought opportunities and challenges to the development of neural networks in the new era.

In the new century, artificial neural networks have achieved rapid development. In 2006, Professor HINTON of the University of Toronto and his doctoral students published a paper in "Science"[9], which triggered an upsurge in the field of research and application of deep learning. Since then, deep learning has been continuously studied in academia.Since 2011, deep learning algorithms have made major breakthroughs in the field of speech recognition research. Speech recognition research experts from Google and Microsoft Research have used deep neural network technology to reduce the error rate of speech recognition by 20% to 30%[10] . In the same year, pharmaceutical companies applied deep learning neural network technology to the problem of drug activity prediction, and achieved the best results in the world [11]. In 2016, Zhang et al. [12] proposed the MS-CNN algorithm, which solved the problem that the first few layers of traditional convolutional neural networks could not be effectively trained through unsupervised prediction training filters on the basis of convolutional neural networks [13]. In terms of recognition, it can effectively recognize visible light natural images and remote sensing images. The application of artificial neural networks has gradually begun to shine in all walks of life. The main application areas are signal processing [14-16], plant diseases and insect pests and irrigation control[17,18], intelligent control of industrial product assembly line [19,20], intelligent driving [21,22], chemical product development [23-25], image processing [26-28], robotic surgery [29-31], automatic control of power systems [32-34], troubleshooting [35,36], process control and optimization[37-39], etc.

## 3 Summary of the structure and development of different neural networks

### 3.1 BP neural network

The network architecture of the BP neural network model (backpropagation algorithm) is multi-layered. It is essentially a gradient descent local optimization technique, which is related to the backward error correction of network weights. The multi-layer structure of the BP neural network makes the output of the model more accurate, but the BP neural network still has certain defects. For non-linear separable problems such as XOR, the use of BP neural network may have a local minimum, which may lead to the inability to find the global optimal solution, and when faced with large sample data, the mean square error MSE is too large and it is difficult to converge. The structure of BP neural network is shown in Figure 1.

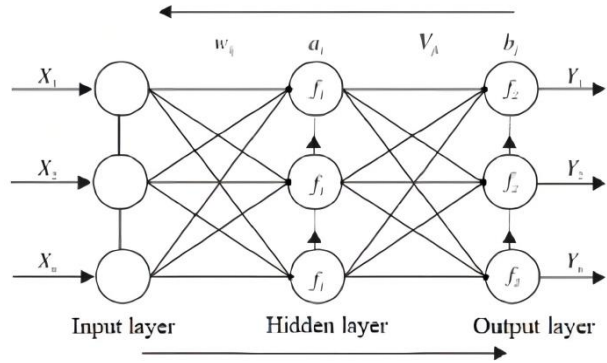

Figure 1    BP neural network structure

Wang Lihong [40] combined the traditional BP to form the AdaBoost-BP model, as shown in Figure 2, the AdaBoost algorithm is trained to calculate the error rate and weight of the first BP model, and the weight is used as the weight parameter of the next BP network. Similar to iterative calculations, where a single traditional BP network hidden layer adopts a 2-layer structure.

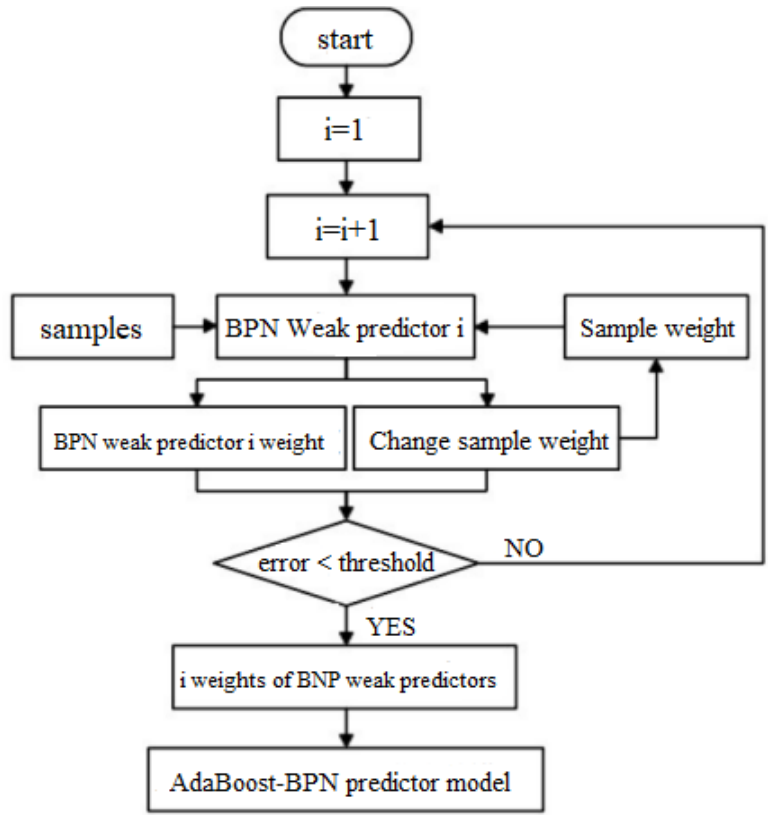

Figure 2    AdaBoost-BP network model flowchart

Aiming at the defect that the BP network uses gradient descent to easily make the model fall into the local optimum, Huang et al. [41] changed the traditional way of BP adjusting its own threshold and weight parameters, and used particle swarm optimization algorithm to obtain the weight and threshold parameters of the BP network.

**3.2 RBF neural network**

RBF Neural Network (Radial Basis Function Neural Network), namely Radial Basis Function Neural Network, proposed by Moody and Darken in 1988, is a local approximation function. For each training sample, only a small amount of weights and thresholds need to be corrected, so the training speed is fast, but it cannot guarantee that the optimal solution will always be obtained [42]. And determining the best structure of the RBF network is not straightforward, it requires trial and error approximation.

RBF is a three-layer forward network with good performance and a single hidden layer. The input layer is composed of signal source nodes, the second layer is the hidden layer, and the third layer is the output layer. The transformation from the input space to the hidden layer space is nonlinear, and the transformation from the hidden layer space to the output layer space is linear, and the transformation function of the hidden unit is the radial basis function. The output layer neuron adopts linear unit, which is a locally distributed non-negative nonlinear function with radial symmetrical attenuation to the center. It has been proved that RBF network is the best approximation of continuous function and is better than BP network [43].

The basic idea of using gradient descent method [44] to calculate neural network weights is: first set error indicators (such as control error indicators, approximation error indicators) according to the control plan, and then use gradient descent method to update the calculated weights according to the error indicators . The control schemes mainly include RBF neural network-based supervisory control, model reference adaptive control and self-tuning control.

**3.3 Convolutional Neural Network**

Convolutional Neural Network (CNN) has achieved good results in image classification and recognition, semantic segmentation, and machine translation. The traditional CNN structure includes four layer structures: convolutional layer, pooling layer, fully connected layer, and output layer. Convolutional neural networks are widely used in the image field and have made great achievements. Especially in image recognition, after a series of operations on convolutional neural networks, the machine can recognize image feature information very accurately. Lou et al. [45] applied VGG16 combined with convolutional neural network CNN to face recognition, and at the same time collected discarded image information and applied it to the original CNN. Compared with the ICA algorithm and the traditional convolutional neural network, the improved model obtained has a significant improvement in performance and image recognition rate. Zhang et al. [46] used CNN for fault detection. The first step of traditional fault detection is to process the signal, and then put the features into the classifier for classification. The intelligent diagnosis and detection algorithm based on convolutional neural network CNN is used to transform the original input signal into two-dimensional image data, and eliminate the interference of experience on feature extraction. Then, the effectiveness of the algorithm is verified by bearing data, and the experiment proves that the algorithm can adapt to changes in workload well.

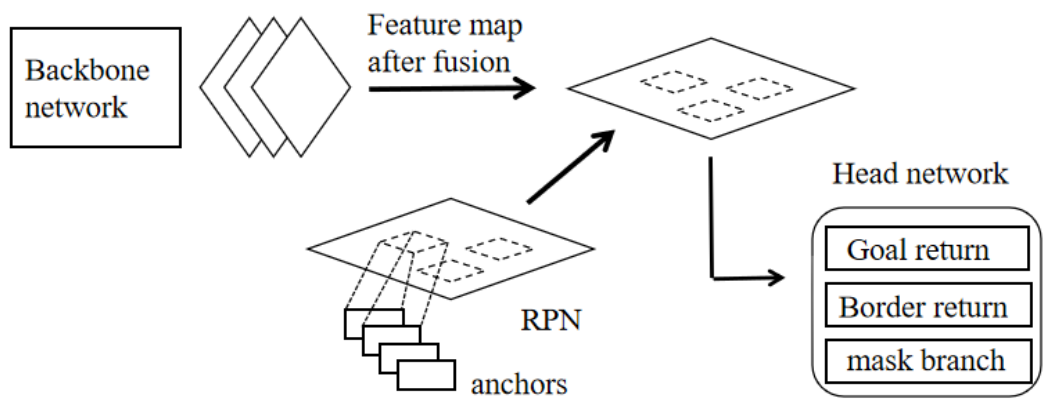

Figure 3   Mask R-CNN model diagram

Mask R-CNN is improved on the basis of Faster R-CNN and is one of the most famous image segmentation models [47]. As shown in Figure 3, the RPN extracts RoIs with inconsistent sizes in the feature map, and then normalizes the RoIs, and uses RoIAlign to replace the RoIPool in Faster R-CNN to ensure the corresponding relationship between input and output.

## 4 Application of neural network in process optimization

With the continuous development of automation processes in various industries, the requirements for automatic control and optimization in chemical processes are getting higher and higher. Due to the uncertainty, nonlinearity, time delay and strong coupling of multiple variables in the chemical control optimization process, the conventional control system is sometimes difficult to work. The emergence of artificial neural networks has further improved the control and optimization results of various processes [47]. The principle of artificial neural network-assisted process control and optimization is: when the process receives a disturbance signal, it is fed back to the artificial neural network, and the neural network revises its mathematical model to control the process system. Here are a few examples of neural networks playing an optimizing role in the chemical process.

### 4.1 Multi-objective optimization of propane recovery process

The condensate recovered by propane is an important chemical raw material and has a wide range of uses. Propane recovery can improve the economic and social benefits of oil and gas field development. Wei et al. [48] established a process model with an improved BP neural network to achieve multi-objective optimization of the recycling process.

### 4.1.1 Improvement of neural network

The output value of the BP neural network directly depends on the weights and thresholds between the neurons in each layer of the link. If the initial weights and thresholds are set unreasonably, the convergence speed of the BP neural network will slow down, and even fall into the local optimal problem [49,50]. The genetic algorithm (GA) has a good global search ability. It screens individuals through selection, mutation, and crossover operations, retains individuals with good fitness values, eliminates individuals with poor fitness, and continues to evolve and iterate to obtain the best individual [51]. The genetic algorithm optimizes the neural network to find the optimal initial weight and threshold of the BP neural network through GA, so that the optimized BP neural network can better predict the output value.

Collect and collect 135 sets of equipment operating parameters of the treatment plant, and substitute them into HYSYS for simulation. Taking the temperature of the cryogenic separator, the temperature of the top of the heavy contact tower, and the reflux temperature of the reflux tank as input, and the output of propane yield and system energy consumption, the first 125 groups are training samples, and the last 10 groups are test samples to establish 3-12-2 BP neural network. The genetic algorithm selects the population number of 100, the number of iterations is 600, the crossover probability is 0.4, and the mutation probability is 0.05. It can be seen from Figure 4 and Figure 5 that the overall trend of the predicted value of yield and energy consumption of the two neural networks is basically the same as the actual output trend. In contrast, the predicted value of GA-BP neural network has higher accuracy. This shows that the improved BP neural network after GA has extremely high accuracy and can be used for subsequent multi-objective optimization models.

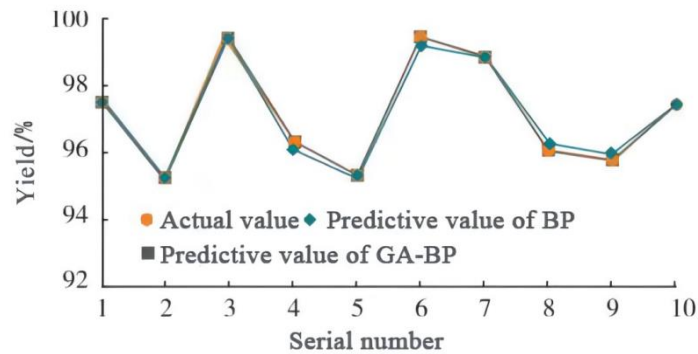

Figure 4   Predicted yield

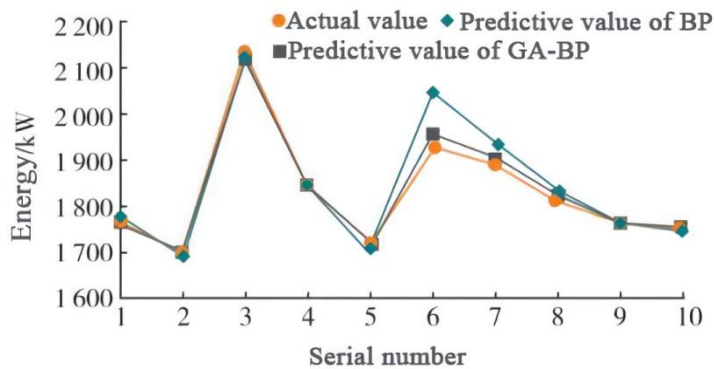

Figure 5   Predicted value of energy consumption

### 4.1.2 Multi-objective optimization

In the multi-objective optimization problem, in most cases, the objectives may conflict with each other, that is, the improvement of one objective will often cause the performance of the other objectives to decrease. At the same time, it is usually impossible to achieve the optimal multiple goals, which makes the multi-objective optimization problem have non-inferior solutions [52]. All non-inferior solutions of multi-objective optimization constitute the Pareto optimal solution set. The output of the trained BP neural network model is used as the objective function of the NSGA-II algorithm. Set the initial population to 100, the number of iterations to 200 generations, and the results obtained by using matlab to calculate are shown in Figure 6.

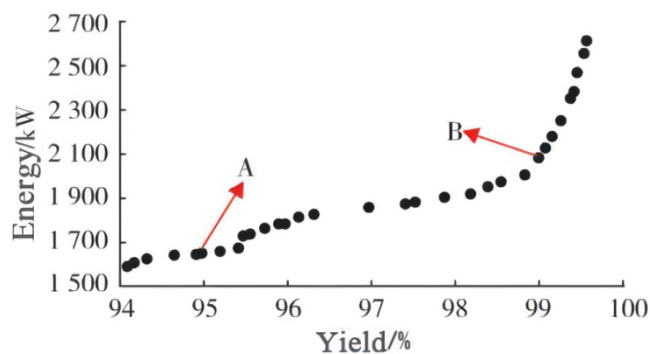

Figure 6    Yield-energy consumption calculation results

By analyzing the key parameters of the DHX process, the following conclusions can be drawn: (1) When the composition of the feed gas is constant, the temperature of the low-temperature separator, the temperature of the heavy contact tower, and the temperature of the reflux tank have a nonlinear effect on the yield and energy consumption of the process. . There is an interaction between the various parameters. (2) The BP neural network improved by GA has a high-precision predictive ability for yield and energy consumption. (3) The relative error between the value in the Pareto solution and the actual output is controlled below 2%, indicating that the Pareto solution is reliable and has certain guiding value for the design of the propane recovery process and the selection of equipment parameters.

## 4.2 Optimization of paraxylene oxidation process

The paraxylene (PX) oxidation device is a major consumer of paraxylene in the entire terephthalic acid (PTA) production process, and it has a great impact on improving the technical and economic indicators of the entire process. Liu [53] used industrial operation data as the basis and used the artificial neural network (BP-ANN) of the error back propagation training algorithm to predict the combustion law of the industrial oxidation reactor. And on this basis, optimization with the goal of reducing consumption has been carried out.

### 4.2.1 Neural Network Design

Select 12 main variables as input layer nodes, tail oxygen and carbon dioxide content as 2 output layer nodes, and establish a three-layer BP network as shown in Figure 6. The transfer function between the input and output of each layer is a Sigmoid nonlinear function, and a deviation constant of 1 is added to the input layer and the hidden layer. In this way, the input vector $U_i$ includes a total of 13 nodes, and in the output vector $S_i$, each component represents the exhaust gas oxygen and carbon dioxide volume content per unit time. The original data required for ANN modeling comes from the production and operation data and analysis data of the industrial PTA oxidation reactor and the first crystallizer of a large chemical company for more than a year. Before using the data, based on the stable regularity of the process operating parameters in a local range, the data is sorted out, some record points that violate the law are eliminated, and the data of each working condition is processed by the average method.

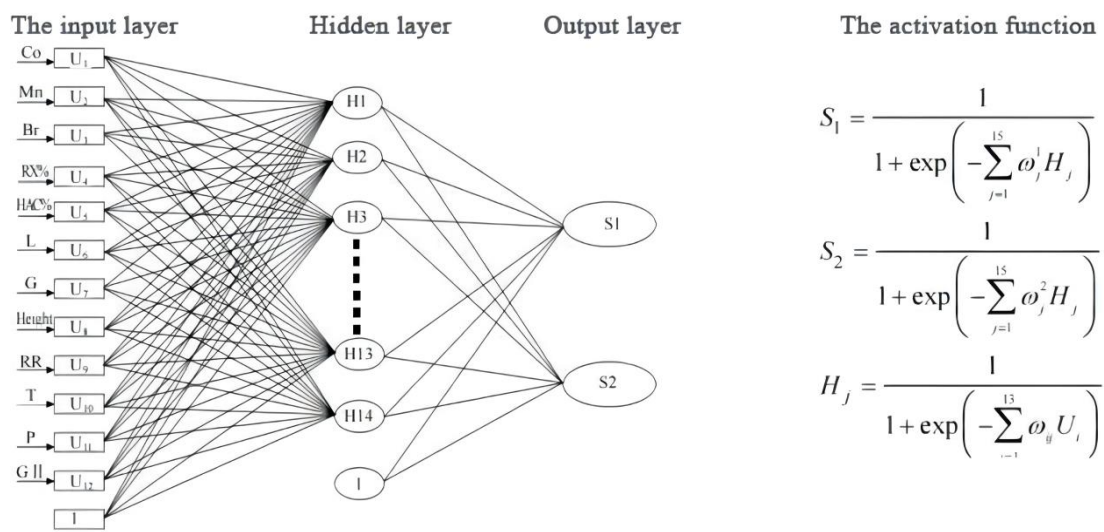

**Figure 7 Neural network structure for paraxylene oxidation model**

**4.2.2 Optimization results**

Figure 7 and Figure 8 reflect the relationship curve of tail gas carbon dioxide calculated by the established prediction model with the catalyst (cobalt) concentration and reflux rate. It can be seen from the figure that when the flow of the reflux liquid during the oxidation process is small, the combustion in the oxidation process is intense and the rate of carbon dioxide generated is higher. With the increase of the reflux, the carbon dioxide first showed a downward trend and then increased. Compared with the mixed liquid in the oxidation reactor, the water concentration in the reflux liquid is higher. Therefore, it also reflects the influence of the water content in the oxidizer on the process. This indicates that there is an optimal value area for adjusting the water content in the reactor through the feed and return flow, and the lowest combustion value can be achieved near it.

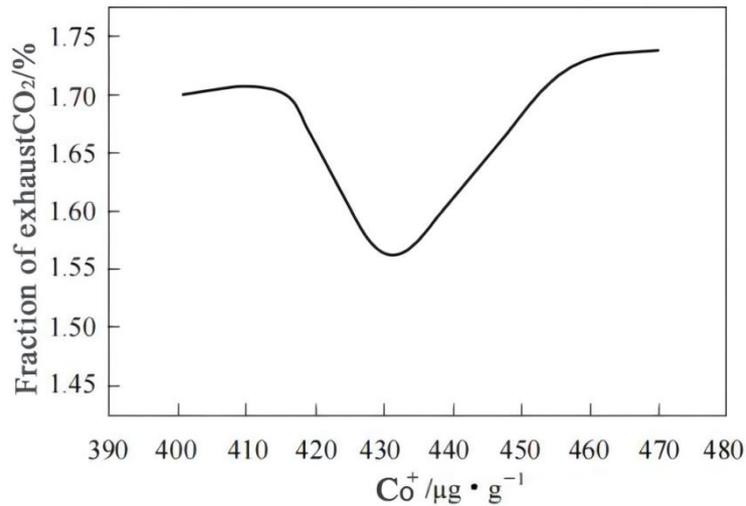

**Figure 8  Relationship between exhaust carbon dioxide concentration and cobalt ion**

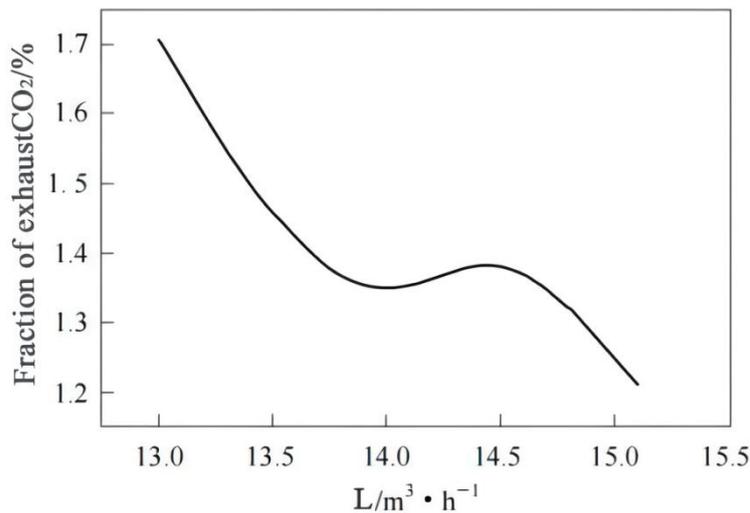

**Figure 9  The relationship between exhaust carbon dioxide concentration and condensing return flow**

Table 1 reflects the existing device operating parameters, optimization results and the results after initial industrial adjustment. It can be seen from the table that the combustion consumption of the paraxylene oxidation process under the existing operating conditions is relatively high. By optimizing the process parameters, the p-xylene processing capacity can be increased by about 1.03%, the oxidation reaction temperature will be reduced by 0.5°C, and the volume fraction of carbon dioxide in the tail gas will be reduced from 1.68% to about 1.23%. Preliminary debugging was carried out with the goal of optimizing operating conditions. The results showed that after adjusting the catalyst concentration and water output in the oxidation process, the quality of the oxidation product was stable, while the volume fraction of carbon dioxide in the tail gas was reduced from 1.68% to 1.35%.

Table 1  Optimization results

| Process conditions | Device status | Optimization Results | Initial adjustment result |
|---|---|---|---|
| Catalyst concentration /μg·g$^{-1}$ | | | |
| Co | 428 | 440 | 420 |
| Mn | 182 | 210 | 200 |
| Br | 550 | 620 | 600 |
| Acetic acid mass fraction /% | 71 | 71 | 71 |
| P-xylene mass fraction /% | 30.8 | 31.5 | 31.5 |
| Liquid flow /m$^3$·h$^{-1}$ | 52.53 | 53.5 | 53.5 |
| Gas flow /m$^3$·h$^{-1}$ | 35319 | 35733 | 35733 |
| Liquid level /% | 79.9 | 78.0 | 78.0 |
| Pumped water flow /m$^3$·h$^{-1}$ | 10.0 | 12.1 | 11.5 |
| Reaction temperature /°C | 196.4 | 195.8 | 195.5 |
| Reaction pressure /MPa | 1.65 | 1.60 | 1.60 |
| Crystallizer air intake /m$^3$·h$^{-1}$ | 2551 | 1930.1 | 1995.0 |
| Exhaust gas composition /% | | | |
| O$_2$ | 4.0 | 3.66 | 3.78 |
| CO$_2$ | 1.68 | 1.23 | 1.35 |

## 4.3  Optimization of the process parameters for the synthesis of geopolymers from fly ash

Geopolymer is a novel aluminosilicate material with an amorphous, three-dimensional structure formed by the reaction of natural aluminosilicate minerals rich in silicon and aluminum components, industrial and mining waste residues and tailings as the main raw materials. It has excellent mechanical properties, fire resistance, high temperature resistance, chemical corrosion resistance, and is widely used [54,55]. Liang et al. [56] used a typical circulating fluidized bed fly ash as a raw material to synthesize geopolymers, and optimized the preparation conditions with the help of BP neural network. Lay a theory for the resource utilization of fly ash in circulating fluidized bed, and point out the direction for the preparation of geopolymers with excellent mechanical properties.

The 6-8-1 three-layer BP neural network structure is used to predict the compressive strength of geopolymers. The training steps are 1,000 steps. The result is shown in Figure 9. It can be seen that the data points drawn by the fitting value of the training sample set and the experimental value are basically distributed on the contour (y=x), indicating that the training result can well map the compressive strength of the geopolymer under different preparation conditions . The average relative error is 0.98%. Only when the error of the training result and the error of the test result are small at the same time, the neural network model obtained has strong generalization ability. It can be seen from the figure that the data points drawn by the fitted value and the experimental value are near the contour, and the relative average error is 3.85%. This shows that the built prediction system has good generalization ability and can be used to predict the compressive strength of geopolymers obtained under different preparation conditions. In order to predict the preparation conditions of the geopolymer with the best mechanical properties, the 6 main factors affecting the compressive strength of the geopolymer were subjected to a 6-factor 5-level orthogonal experiment, with a total of 25 combinations. The modulus of the activator is 0.8 ~ 1.6, the liquid-solid ratio is 0.7 ~ 0.9, the mass fraction of $Na_2O$ is 0.07 ~ 0.15, the curing temperature is 20 ~ 100 °C, the curing time is 8 ~ 24 h, and the stirring time is 2 ~ 30 min. The 25 combinations determined by the orthogonal experiment were used as the input sample set and normalized.

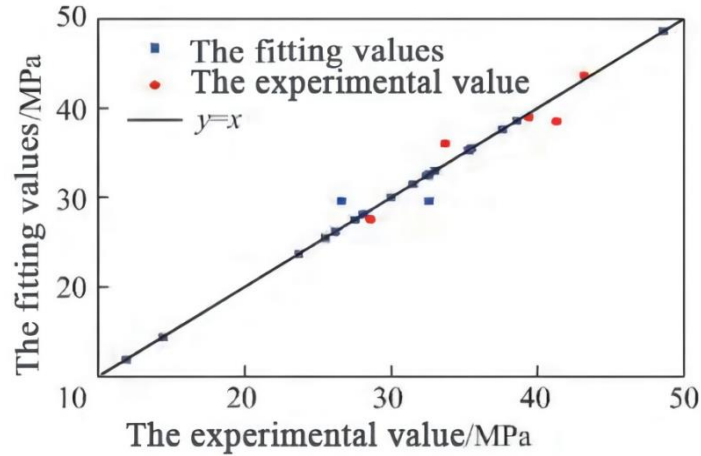

**Figure 10　Neural network prediction**

**results of compressive strength of geopolymers**

Refer to the trained BP neural network prediction system for simulation prediction. Taking the experimental sample cured for 7 days at room temperature after demolding as an example, the preparation conditions of the geopolymer with the best mechanical properties are: activator modulus 1.6, liquid-solid ratio 0.8, $Na_2O$ mass fraction 9%; curing temperature 20 ℃; curing time 24h, and the stirring time 20 min.

## 5 Conclusion

In summary, artificial neural networks have been widely used in chemical process optimization and are gradually becoming an indispensable tool in the chemical process. They play an irreplaceable role in many fields, saving a lot of manpower, material and financial resources and can get more perfect results. Of course, people's understanding and research on artificial neural networks need to be improved. Nowadays, most people still use simpler network models, and there is still a lot of room for development. At present, various fields are developing in the direction of artificial intelligence, constantly enriching and accumulating theoretical levels, and improving the technical reliability of artificial neural networks. And try to develop intelligent optimization control simulation software based on artificial neural network, which will play a vital role in the development of my country's chemical industry in the future.